# Few Layer $HfS_2$ FET

**Toru Kanazawa[1*], Tomohiro Amemiya[2,3*], Atsushi Ishikawa[3,4], Vikrant Upadhyaya[1], Kenji Tsuruta[4], Takuo Tanaka[3], and Yasuyuki Miyamoto[1]**

[1]Department of Physical Electronics, Tokyo Institute of Technology, 2260002 Japan

[2]Quantum Nano Electronics Research Center, Tokyo Institute of Technology, 2260002 Japan

[3]RIKEN, 3510198, Japan

[4]Department of Electrical & Electronic Engineering, Okayama University, 7008530, Japan

*e-mail: kanazawa.t.aa@m.titech.ac.jp, amemiya.t.ab@m.titech.ac.jp

**2D materials are expected to be favorable channel materials for field-effect transistor (FET) with extremely short channel length because of their superior immunity to short-channel effects (SCE). Graphene[1,2], which is the most famous 2D material, has no bandgap without additional techniques[3,4] and this property is major hindrance in reducing the drain leakage. Therefore, 2D materials with finite band gap, such as transition metal dichalcogenides (TMDs, e.g. $MoS_2$,[5,6] $WSe_2$[7]) or phosphorene[8,9], are required for the low power consumption FETs. Hafnium disulfide ($HfS_2$) is a novel TMD, which has not been investigated as channel material. We focused on its potential for well-balanced mobility[10] and bandgap[11] properties. The higher electron affinity of Hf dichalcogenides compared with Mo or W chalcogenides facilitates the formation of low resistance contact and staggered heterojunction with other 2D materials. Here we demonstrate the first few layer $HfS_2$ FET with robust current saturation and high current on/off ratio of more than $10^4$.**

In order to realize ultra-low power circuits, three important electrical characteristics are desirable in channel material for FETs. They are low supply voltage, high drivability and small off-leakage current. The former two are achievable by channel scaling in the same manner as before. On the other hand, continuous scaling of channel length $L_{ch}$ has led to severe SCE. SCE increases the off-leakage current and standby power consumption of logic circuits. One of the possible solution to reduce the SCE in ultra-scaled channel ($L_{ch}<10$ nm) is the introduction of atomically thin body channel structure. However, in extremely thin body channel consisting of conventional semiconductors such as Silicon[12], InGaAs[13] and Germanium[14], surface roughness scattering severely reduces the carrier mobility[15] and ballistic coefficient. 2D materials have layered crystal structure with strong covalent/ionic bond in plane and weak Van der Waals interaction between layers. As a result of this unique property, 2D materials empower in obtaining the atomically flat surface with discrete thickness defined by its single-layer thickness and carrier transport with minimal surface roughness scattering. These features of 2D material can be useful in obtaining ultra-low power operation because of the attainment of high mobility with the extremely short and thin channel design. Another critical property for low leakage current operation is the value of band gap. Zero (or very small) bandgap materials such as graphene show ambipolar behavior in field-effect current modulation. The off current could not be small enough owing to reverse polarity operation. Although, this property is useful for many applications (e.g. RF applications[16], sensing[17]), it becomes unpalatable for low-power logic circuits[18]. Moreover, to reduce the drain leakage current owing to band-to-band tunneling such as GIDL (gate induced drain leakage)[19], wide bandgap is desirable for materials with finite bandgap. Therefore, the mobility and bandgap should be carefully and widely investigated for various applications and performance improvements.

$HfS_2$ was previously supposed to be semi-insulating material because the measured conductivity of $HfS_2$ was very low compared with $MoS_2$[20,21]. There were few reports which mainly investigated on electron transport and field-effect doping properties of $HfS_2$. However, in recent reports, some interesting properties of single layered $HfS_2$ have been theoretically predicted. The long-wave acoustic phonon limited mobility, $\mu_{AP}$ of TMDs estimated in ref. 10 and energy bandgap $E_g$ in ref. 11 are plotted in Fig. 1. It indicates that there is substantial trade-off between $\mu_{AP}$ and $E_g$. Single layer $HfS_2$ is expected to have good upper limit of mobility (~1800 $cm^2$/Vs) and reasonable energy band gap (~1.2 eV) for high on/off ratio. Although several reports concerning the electron mobility in $MoS_2$[6,22,23] suggested the existence of several other scattering mechanisms, we comprehended that the comparison of mobilities in TMDs by the uniform calculation method will aptly represent relative trend among them. Consequently, we planned to experimentally evaluate the FET performance of $HfS_2$ and reveal its possibility for transistor channel. Also, the other properties of thin layered HfS2 have not been investigated thoroughly and we would report some basic characteristics of atomically thin-layered $HfS_2$ essential for fabrication and characterization.

Figure 2 shows (a) a piece of $HfS_2$ single crystal which is commercially available with high orientation and purity. We performed the mechanical exfoliation using scotch tape to obtain the thin-film $HfS_2$ on the substrate. First, a small piece of $HfS_2$ was spread on tape and then cleaved several times so as to reduce it to the average thickness of layers as shown in Fig. 2 (b). Thickness dependent color contrast was distinctly observable. Figure 2 (c) shows the optical image of exfoliated thin-film $HfS_2$ on 285-nm-thick $SiO_2$/Si substrate. Triangular or hexagonal shape often appeared in the flakes. These cleaved edges indicate the crystal orientation of exfoliated $HfS_2$. Incidentally, hexamethyldisilazane (HMDS) treatment supports in peeling a large size (>10 μm) and atomically thin flake. The Raman spectrum of thin $HfS_2$ layers (<10 nm) on $SiO_2$/Si substrate with excitation wavelength of 532 nm is shown in Fig. 2 (d). The primary peak appeared at the Raman shift of around 337 cm$^{-1}$ and it was consistent with the previous experimental report of bulk $HfS_2$ depicting first order $A_{1g}$ peak[24]. Satellite peaks at 260 and 321 cm$^{-1}$ were also considered as $E_g$ and $E_g$ (LO). These results indicate that single crystal layers with well aligned atoms remained intact during exfoliation and organic solvent cleaning. At this time, there was no remarkable change from bulk to thin films.

$HfS_2$ has a $CdI_2$-like octahedral coordinate layered structure[24] with atomic layer thickness of 0.59 nm[25] (Fig. 3 (a)). Hf atoms (blue spheres) are sandwiched by S atoms (yellow spheres) and Hf atoms in each plane were stacked at same position (1T, tetragonal symmetry). This crystal structure is different from $MoS_2$ which typically have trigonal prismatic coordinate with 2H symmetry. It is anticipated that the electron effective mass at conduction band minima of $HfS_2$ and other octahedral coordinate TMDs have strong anisotropy and hence orientation-dependent mobility might be observed. The schematic image of fabricated FET with proper biases is shown in Fig. 3 (b). The thin body channel contains several atomic layers which were transferred on to the atomic layer deposited (ALD) 75-nm-thick $Al_2O_3$/$p^{++}$-Si(100) substrate. We used plasma enhanced chemical vapor deposited $SiO_2$ (285 nm) as back-gate insulator. However, it could not work with clear current modulation by back gate because of the small gate capacitance and non-uniformity of $SiO_2$ film quality (see supplementary Fig. S3). The visibility of atomically thin $HfS_2$ layers might be strongly dependent on the thickness and permittivity of insulator as observed for graphene[26] and $MoS_2$[27]. In our structure, the contrast of single layer $HfS_2$ was too weak to form electrodes with the alignment technique. However, several layers of $HfS_2$ were clearly observable by optical microscope and atomic force microscope (AFM) (see supplementary Fig. S4). Au/Ti electrodes were fabricated on thin-film $HfS_2$ as source and drain contacts. Source electrode was connected to the ground and drain electrode was positively biased with respect to the source ($V_{DS}$). Back-gate voltage $V_{GS}$ was applied to the $p^{++}$-Si substrate. Optical microscope measurement suggested the channel width and length of 10 μm (at the source edge) and 2 μm, respectively (Fig. 3 (c)). According to the cross-sectional height profile measured by AFM in Fig. 3 (d), thickness of the $HfS_2$ channel was around 3.8 nm which is composed of 6 atomic layers.

Figure 4 (a) indicate the output characteristics of fabricated thin-film $HfS_2$ FETs at room temperature.

The clear saturation behavior was observed in $I_D$-$V_{DS}$ curves for all $V_{GS}$ biases and sweep of $V_{DS}$ (0~5 V). This robust current saturation at high $V_{DS}$ (5 V) indicates that the bandgap of $HfS_2$ could be sufficient for short channel FET fabrication. $I_D$ increments consecutively over range of measured $V_{GS}$. Unfortunately, our measurement system is not able to apply voltage over 40 V. Fully n-type enhancement mode operation with threshold voltage greater than 8V was observed. $I_D$-$V_{DS}$ curves at small bias regime show linearity without offset bias. Although instability of I-V curves depend on measurement cycles and sweep conditions, the current modulations were constantly obtained for over a month in the atmosphere. A thin $HfS_2$ has a good tolerance against moisture and oxygen. $I_D$-$V_{GS}$ (transfer) characteristics with logarithmic (left-hand side) and linear (right-hand side) scale for $V_{DS}$ of 3 V are also shown in Fig. 4 (b). The maximum drain current of 0.2 μA/μm was obtained at $V_{GS}$ = 40 V. The on/off ratio was over 10,000 when the $V_{GS}$ was varied from −5 V (off state) to 40 V (on state). This device had large hysteresis (~15 V) (see supplementary Fig. S5). It could have been caused by response of traps at the interface between $HfS_2$ and $Al_2O_3$ or in the bulk $Al_2O_3$ because top-gate FET with electron double layer (EDL) structure did not show large hysteresis (also see in Fig. S6). The gate leakage current $I_G$ is also plotted in logarithmic scale and it indicates that $I_G$ will be comparable to $I_D$ for higher values of $V_{GS}$. Therefore, the on current and other performances of the present device can be improved by an efficient modulation of the surface potential using the thinner gate dielectric with good interface properties.

In conclusion, we demonstrated the fabrication and I-V characteristics of $HfS_2$ FETs. Mechanical exfoliation by scotch tape provides atomically thin $HfS_2$ single crystal layer. For the channel thickness of 3.8 nm, robust saturation behavior and drain current of 0.2 μA/μm were observed with high on/off current ratio ($> 10^4$). These results provided basic knowledge of $HfS_2$ as a channel material for FET and the attractive properties deemed to be significant for ultra-low power applications were experimentally demonstrated.

**Methods:**

At first, 75-nm-thick $Al_2O_3$ was deposited on $p^{++}$-silicon (ρ<0.001 Ωcm) by thermal atomic layer deposition (ALD) system (Ultratech/Chambridge Nanotech Savannah S100) as a back-gate insulator. Next, $HfS_2$ flakes were mechanically exfoliated from the highly oriented crystal by scotch tape method and transferred on $Al_2O_3$ surface. The $Al_2O_3$ surface was passivated by HMDS to prepare a hydrophobic surface which is suitable for bonding with $HfS_2$ see supplementary S2. In this report, exfoliated $HfS_2$ flakes were not intentionally doped for both the channel and contact regions. After the optical observation of transferred films, the alignment exposure for source and drain electrodes was carried out by EB lithography (Crestec CABL-9000) using PMMA (polymethyl methacrylate). Then, Ti(20 nm)/Au(100 nm) were EB evaporated and lifted-off. Finally back-gate contact was formed by EB evaporation of Cr (20 nm) and Au (100 nm).

Thickness of $HfS_2$ thin films were evaluated by AFM (Veeco Nanoscope III). All DC characteristics reported in this letter were measured by Agilent 4155B semiconductor parameter analyzer.

**Acknowledgements**

The authors thank S. Tamura for technical support with electron-beam lithography.

**Authors Contribution**

T.K., T.A., A.I. and Y.M. conceived and designed the experiments. T.K. and V.U. fabricated the samples. T.K. and T.A. carried out the DC measurement. A.I. and T.T. performed the total-reflection Raman spectroscopy measurement. T.K., A.I., V.U. and T.T. carried out the AFM measurement. A.I. and K.T. contributed to the electrolyte-gate device fabrication. T.K., T.A., A.I., V.U. and Y.M. wrote the paper. T.A. and Y.M. organize the research group. All authors discussed the results and commented on the manuscript.

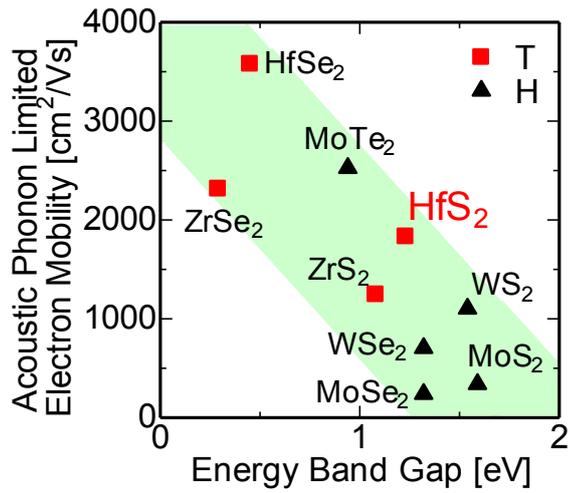

**Figure 1 | Mobility and Bandgap plot of TMDs.** Acoustic phonon limited electron mobilities and energy band gap calculated in references 10 and 11 are plotted in the graph. Patterns indicate the coordinate structure (T: octahedral coordinate and H: triangle prism coordinate).

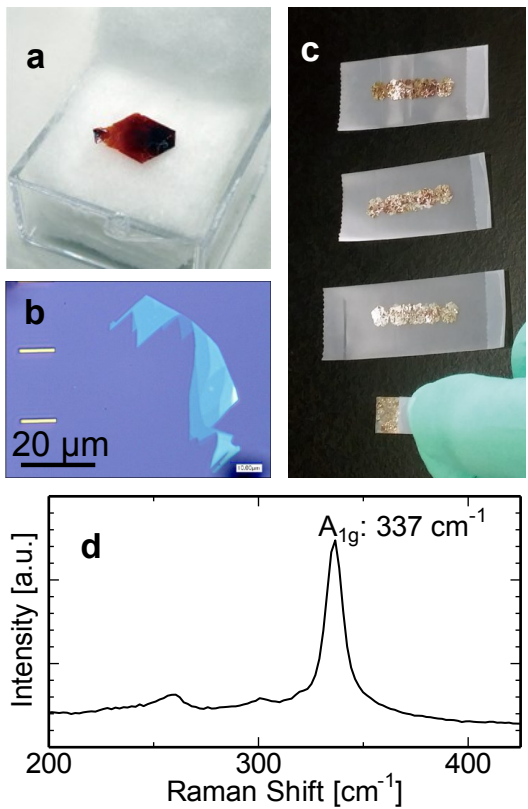

**Figure 2 | Optical image of HfS$_2$. a**, single crystal piece of HfS$_2$. **b**, mechanical cleavage of HfS$_2$ using the "Scotch tape" method. Upper images show thicker layers, which were thinned by continuous exfoliations. **c**, HfS$_2$ transferred on 285-nm-thick SiO$_2$ deposited on Silicon substrate. **d**, Raman spectrum of thin-film HfS$_2$ on SiO$_2$/Si substrate. Primary A$_{1g}$ peak was observed at 337 cm$^{-1}$ with some satellite peaks.

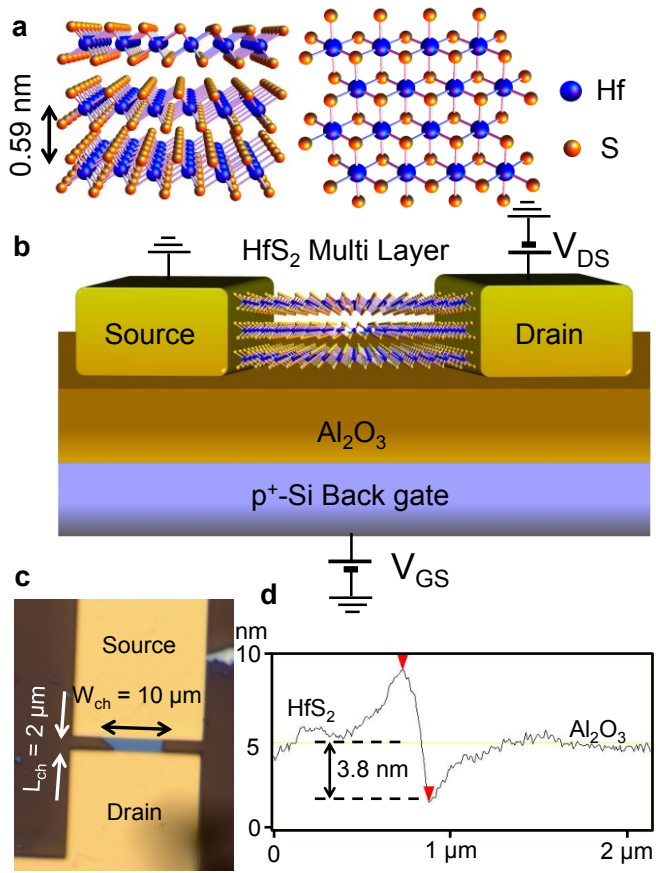

**Figure 3 | Device structure of fabricated HfS$_2$ FET. a**, schematic of the device structure with crystal structure. 3.6-nm-thick HfS$_2$ layers were exfoliated on 75-nm-thick Al$_2$O$_3$, which was atomic layer deposited on degenerately doped p+-Si substrate. The source electrode was grounded and gate bias was applied through the Si substrate as a back gate electrode. Channel direction was designed along with <1200> orientation. Thickness of an atomic layer is considered to be around 0.59 nm. **b**, Optical image of the fabricated HfS$_2$ FET device. The channel length and width at source edge were estimated to be around 2 and 10 μm respectively. **c**, Height profile of channel layer obtained by AFM. The thickness was around 3.6 nm and it suggested that the channel contained about 6 atomic layers of HfS$_2$.

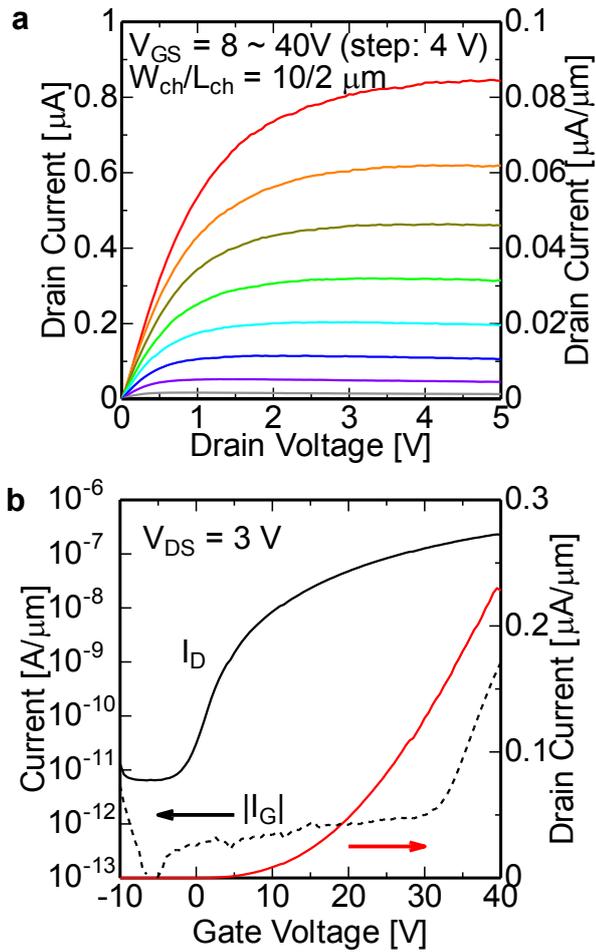

**Figure 4 | DC characteristics of thin-layered HfS$_2$ FET. a**, Output characteristics with channel width and length of 10 and 2 μm at room temperature respectively. Current modulation property and robust saturation behavior were observed. **b**, Transfer characteristics with V$_{ds}$ of 3 V. The maximum drain current obtained in this device was 0.2 μA/μm. The minimum drain current was less than 10 pA/μm and hence the on/off current ratio in this voltage condition is over 10$^4$. Gate leakage current is smaller than drain current in whole bias range but not negligible under strong field. The FET was operated as enhancement mode device.

**Supplementary Information for "A few layered HfS$_2$ FET"**

**Thickness and contrast evaluation of HfS$_2$ layers**

We confirmed the layered exfoliation of HfS$_2$ on SiO$_2$ substrate. The optical image has clear discrete color variation from blue (thin) to white (thick). CCD count rate along the white dotted line from A to B is also described. The counting values were decreased in incremental steps from substrate to thicker HfS$_2$.

AFM measurement showed triangular structure and the smallest step size was estimated to be less than 0.9 nm. It is reasonable to consider monolayer HfS$_2$ equal to step thickness

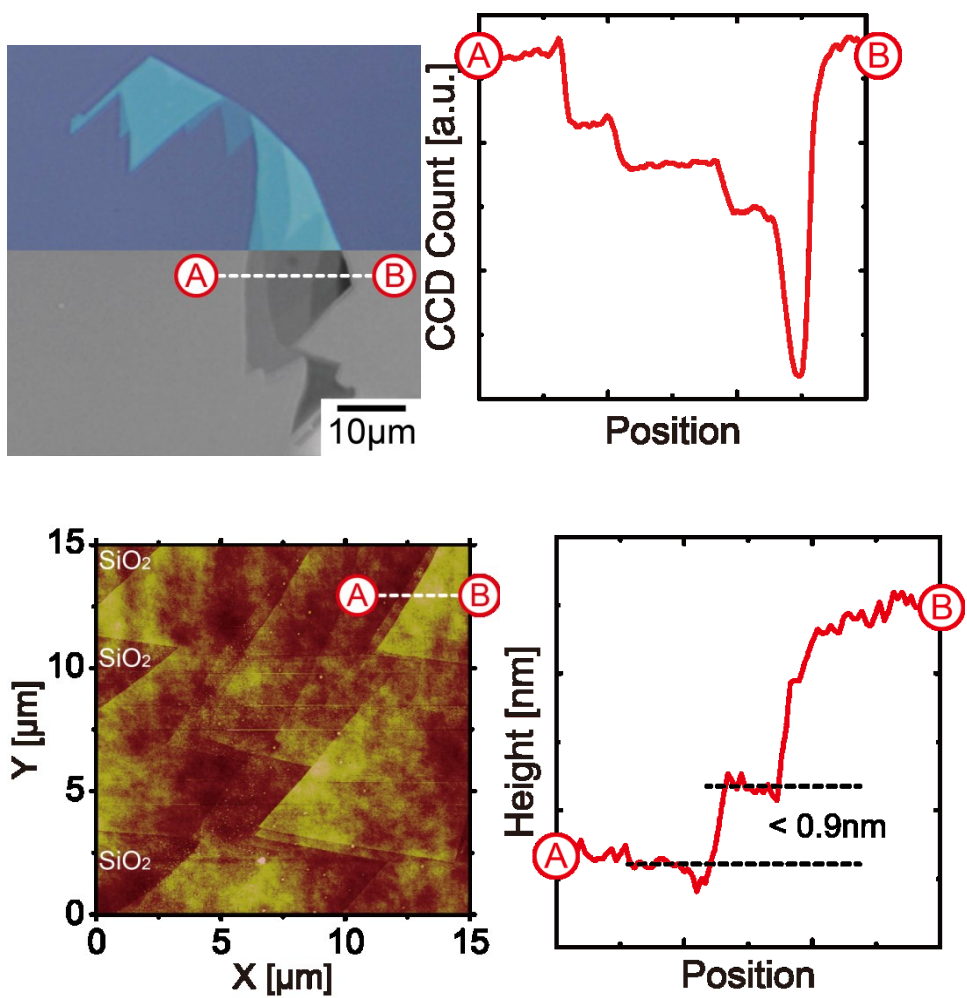

**Figure S1 | Optical contrast and ML thickness of HfS$_2$/SiO$_2$ 285 nm/Si. a**, Optical Image and contrast profile **b**, AFM image and cut line profile.

**Hydrophyte evaluation of HfS$_2$**

The contact angles of drops of water were evaluated for HfS$_2$, highly oriented pyrolytic graphene, SiO$_2$, O$_3$ cleaned SiO$_2$ and HMDS-SAM treated SiO$_2$. HfS$_2$ crystal surface showed strong hydrophobicity similar to graphene. It means that HMDS SAM treatment is suitable to transfer not only graphene but also HfS$_2$.

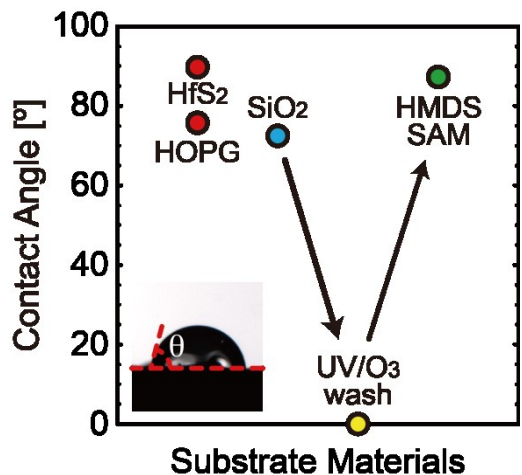

**Figure S2 | Contact angle evaluation of HfS$_2$, graphene, SiO$_2$ and HMDS SAM treated SiO$_2$**

**Current properties of HfS$_2$ FET with SiO$_2$/Si back gate.**

The current properties of HfS$_2$ FET with SiO$_2$/Si back gate operation are shown below. The on/off ratio is only 10 for the bias range of 0~20 V. This inferior modulation properties by $V_{GS}$ is owing to the defect traps and non-uniformity of SiO$_2$ back-gate insulator deposited by plasma enhanced chemical vapor deposition.

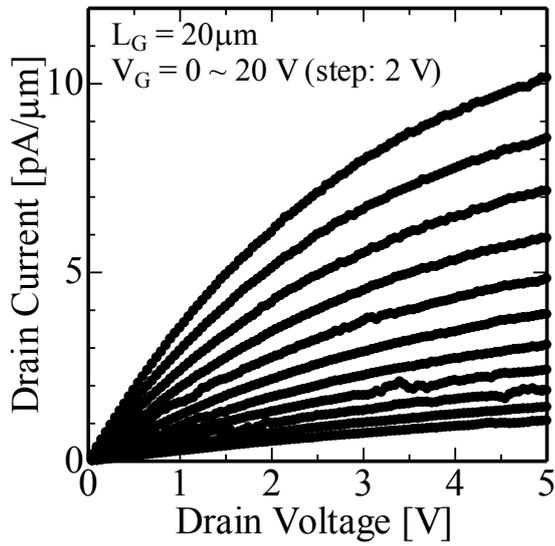

**Fig. S3 | $I_D$-$V_{DS}$ characteristics of the HfS2 FET formed with (SiO$_2$)/Si back gate.**

**Thickness and Contrast evaluation of HfS$_2$ on Al$_2$O$_3$**

Although, the contrast of single step layer against substrate was very small, few layers of HfS$_2$ were easily observed by optical microscope. These steps indicate the thickness which correspond to 1~2 monolayers of HfS$_2$.

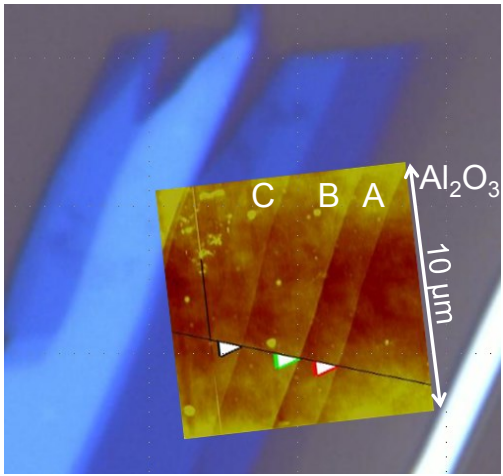

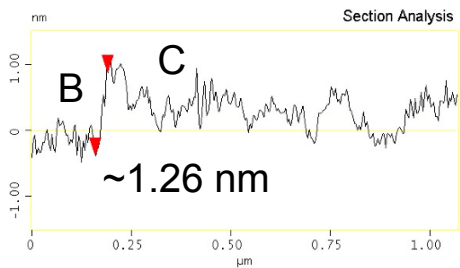

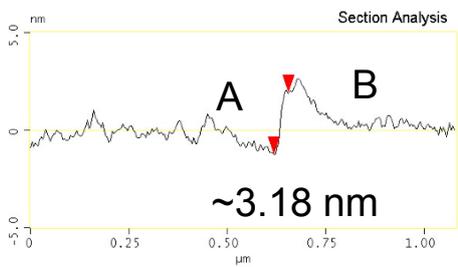

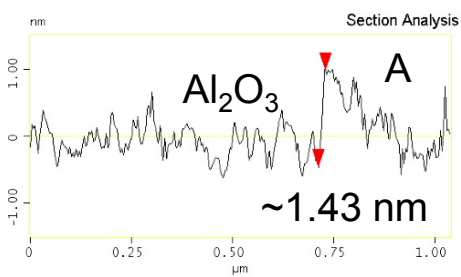

**Figure S4 | Optical image and AFM profiles of HfS$_2$/Al$_2$O$_3$ 75 nm/Si.**

**Hysteresis characteristics of HfS$_2$ MOSFET**

Here, we discuss about hysteresis characteristics of HfS$_2$ FET for back-gated operation. Fig. S4 shows the logarithmic plot of I$_D$-V$_{GS}$ characteristics. The ΔV$_{GS}$ shift in subthreshold region in hysteresis characteristics was around 15 V.

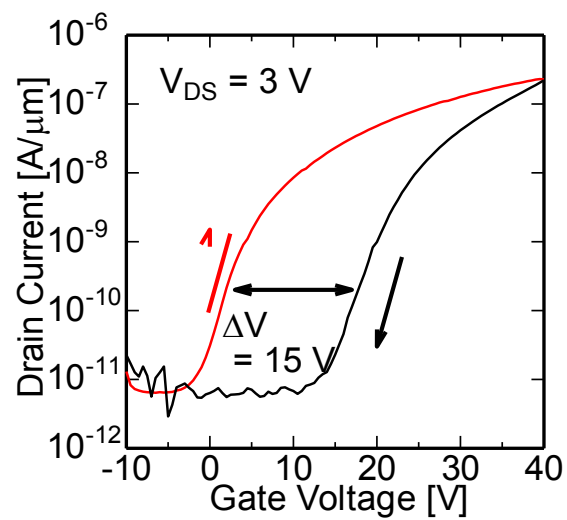

**Figure S5 | Double-sweep transfer characteristics of HfS$_2$ transistor.**

## Current characteristics of electron double layer (EDL) transistor

Figure S6 indicates the DC characteristics of electron double layer FET with $HfS_2$ channel and $PEO:LiClO_4$ as the gate dielectric (as schematically shown in **a**). The maximum drain current of EDL FET reached 750 μA/μm. The drain current was 1000 times larger than back-gate operation. Hysteresis characteristics and on/off ratio were also improved in EDL transistors. There exists small hysteresis and gate current was probably dominated by response time of electrolyte

The effective mobility was extracted from transfer characteristics for $V_{DS}$ = 100 mV using equation $\mu_{eff} = \frac{\partial I_D}{\partial V_{GS}} \frac{L}{W C_{OX}} \frac{1}{V_{DS}}$. EDL thickness was not certain and assumed to be 1~5 nm. Then, the estimated value of $\mu_{eff}$ was 17~85 cm$^2$/Vs. The effective $V_{GS}$ seemed to be smaller than applied $V_{GS}$ because of the small gate contact area and long distance between gate and channel (100 μm~). However, these results indicate the possibility of high current performance using $HfS_2$ as a channel material with high on/off current ratio.

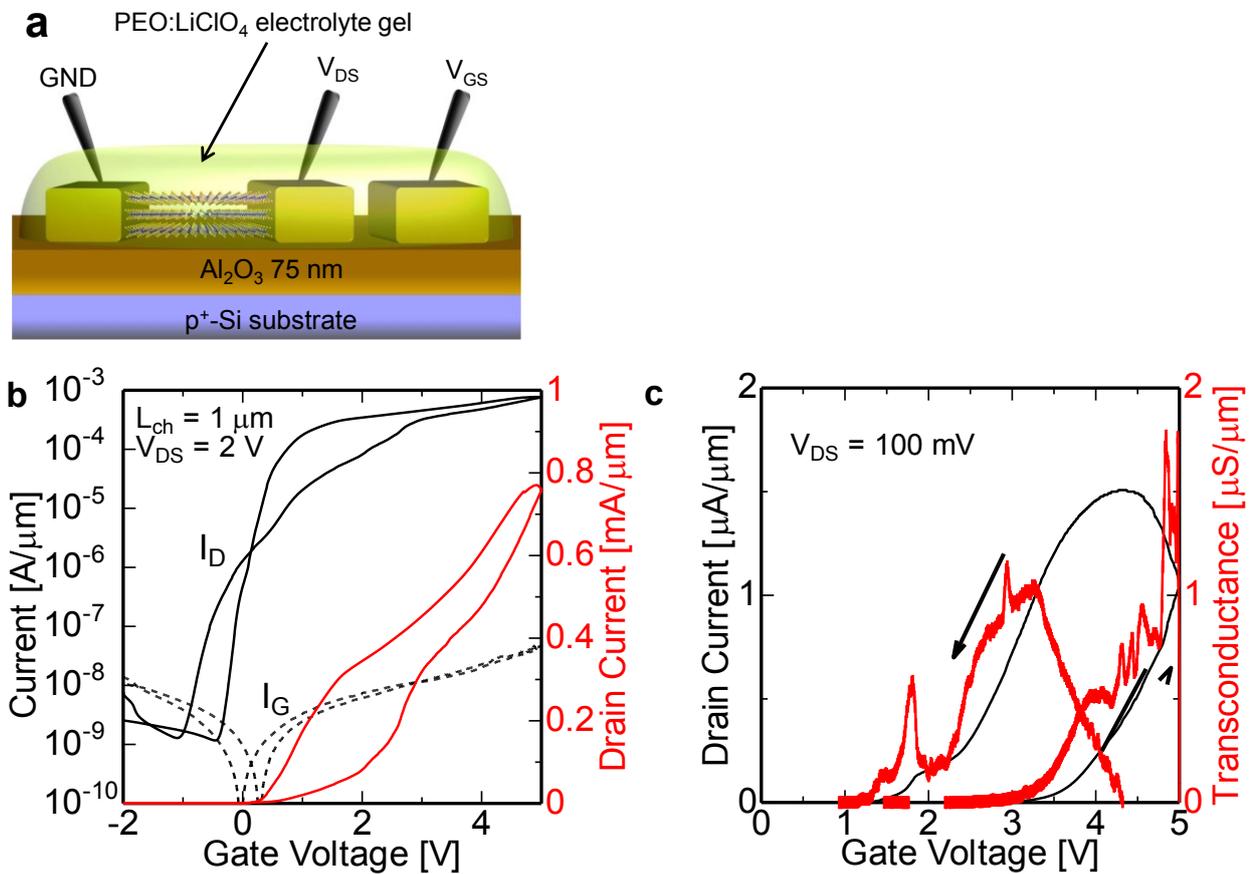

**Figure S6 | I-V characteristics of EDL transistor. a**, Schematic image of fabricated EDL transistor. **b**, Transfer characteristics at $V_{DS}$ = 2 V and **c**, at $V_{DS}$ = 100 mV.